\documentclass[prd,aps,showpacs,nofootinbib,superscriptaddress]{revtex4}
\usepackage{epsfig}
\usepackage{amsmath, slashed}
\usepackage{xcolor}
\usepackage{dsfont}

\newcommand{\uvec}{\boldsymbol}
\newcommand{\ud}{\mathrm{d}}

\allowdisplaybreaks
\begin{document}

\title{On the deuteron relativistic charge distributions}

\author{C\'edric Lorc\'e}
\affiliation{CPHT, CNRS, Ecole Polytechnique, Institut Polytechnique de Paris, Route de Saclay, 91128 Palaiseau, France}

\author{Pierre Wang}
\affiliation{ENSTA Paris, Institut Polytechnique de Paris, Route de Saclay, 91120 Palaiseau, France}

\begin{abstract}
We study the relativistic 2D charge distributions in the case of a spin-$1$ target. These charge distributions are based on a phase-space approach allowing one to study their frame dependence, and hence to relate the familiar rest-frame picture with the light-front picture developed in the last two decades. Like in the spin-$1/2$ case, we show that relativistic kinematical effects associated with spin are responsible for the distortions of the charge distributions seen in a moving frame. Applying our results to the deuteron, we observe a mild frame dependence compared to the nucleon case.
\end{abstract}



\maketitle

\section{Introduction}

Over the last decades, the electromagnetic form factors of hadrons have been measured with impressive precision, see e.g.~\cite{Arrington:2006zm,Perdrisat:2006hj,Punjabi:2015bba,Pacetti:2015iqa}. Since they describe elastic scattering off a target, they encode key information about the spatial distribution of electric charge and magnetization. In order to define the latter, one usually considers Fourier transforms of the form factors in the Breit frame~\cite{Ernst:1960zza,Sachs:1962zzc}, where the spin structure resembles that of the non-relativistic theory. Unfortunately, the relativistic framework appears to be incompatible with a probabilistic picture in the target rest frame, and so the interpretation of the Breit-frame distributions is often thought to be plagued by unclear relativistic corrections~\cite{Yennie:1957,Breit:1964ga,Kelly:2002if}.

The only way to avoid relativistic corrections in this picture is to consider that the system is moving with almost the speed of light relative to the observer. This can be achieved either by working in the infinite-momentum frame or by using the light-front formalism~\cite{Fleming:1974af,Soper:1976jc,Burkardt:2000za,Miller:2010nz}. Besides losing one spatial dimension, the drawback is that these charge distributions get distorted by effects associated with the motion of the target~\cite{Burkardt:2002hr,Carlson:2007xd} and sometimes seem hard to reconcile with the rest-frame picture~\cite{Miller:2007uy}.

Since a strict probabilistic interpretation requires Galilean symmetry and since relativistic spatial distributions are necessarily frame-dependent, we recently proposed to change the perspective by adopting a phase-space approach and contenting ourselves with a quasi-probabilistic but fully relativistic picture~\cite{Lorce:2017wkb,Lorce:2018egm}. Once properly normalized, Breit-frame distributions are then reinstated as meaningful quantities representing physical properties of the system in the (average) rest frame. Under a Lorentz boost and after integration over the longitudinal spatial coordinate, these new relativistic distributions show how the rest-frame picture gets distorted by kinematical effects associated with the spin of the target~\cite{Lorce:2020onh}. The phase-space approach therefore allows one to connect in a smooth way the Breit-frame distributions to the light-front ones. 

In Ref.~\cite{Lorce:2020onh} we studied in detail the structure and frame dependence of relativistic 2D charge distributions for a spin-$1/2$ target, and then applied the formalism to the nucleon whose electromagnetic form factors are well measured. The aim of the present work is to extend this study to the case of a spin-$1$ target. The reason is that, as the spin of the target increases, new features associated with higher multipoles appear and lead to a deeper understanding of the underlying physics. Although the focus of this work is mostly theoretical, it is also of experimental interest since it clarifies the model-independent physical interpretation of the deuteron\footnote{Our formalism applies naturally also to the $\rho$ meson, but the corresponding electromagnetic form factors have not yet been measured.} electromagnetic form factors. The unpolarized deuteron form factors $A(Q^2)$ and $B(Q^2)$ are pretty well constrained experimentally, but a better measurement of the elastic tensor analyzing power $T_{20}(Q^2)$ is needed for a more reliable separation into the usual Breit-frame multipole form factors~\cite{Holt:2012gg}. One of the goals of the C1-approved Jefferson Lab experiment C12-15-005~\cite{Long:2015} is to improve the current situation. There are also prospects for measuring deuteron electromagnetic form factors over a larger $Q^2$-range at the future Electron-Ion Collider (EIC)~\cite{AbdulKhalek:2021gbh}.

The rest of this paper is organized as follows. In Section~\ref{sec:2} we quickly review the phase-space approach and define relativistic 2D distributions for arbitrary average momentum of the target. Then we remind in Section~\ref{sec:3} the parametrization of spin-$1$ matrix elements of the charge current operator along with the definition of multipole form factors. Relativistic distortions of the amplitudes induced by the motion of the target are discussed in Section~\ref{sec:4}, and our results for a spin-$1$ target are presented in Section~\ref{sec:5}. We then perform the Fourier transforms in Section~\ref{sec:6} and show the relativistic charge distributions derived from the phenomenological deuteron form factors for different values of the deuteron average momentum. Finally we summarize our findings in Section~\ref{sec:7}.

\section{Phase-space approach}\label{sec:2}

In order to define the internal charge distribution of a massive system, we first need to define the position of the latter. In the relativistic context we must also in general specify the frame in which this position is defined, and hence specify the momentum of the system. As a result, we are naturally led to the phase-space perspective.

It has been shown long ago that the expectation value of an operator $O$ in a physical state $|\Psi\rangle$ can be written in the convenient form~\cite{Wigner:1932eb,Hillery:1983ms}
\begin{equation}\label{PSampl}
\langle O\rangle_\Psi=\int\frac{\ud^3P}{(2\pi)^3}\,\ud^3R\,\rho_\Psi(\uvec R,\uvec P)\,\langle O\rangle_{\uvec R,\uvec P}.
\end{equation}
The information about the global wave packet of the system
\begin{equation}
    \tilde\Psi(\uvec p)=\frac{\langle p|\Psi\rangle}{\sqrt{2E_p}},\qquad \Psi(\uvec x)=\int\frac{\ud^3p}{(2\pi)^3}\,e^{i\uvec p\cdot\uvec x}\,\tilde\Psi(\uvec p),
\end{equation}
where $|p\rangle$ is a four-momentum eigenstate normalized as $\langle p'|p\rangle=2E_p\,(2\pi)^3\delta^{(3)}(\uvec p'-\uvec p)$ and $E_p=\sqrt{\uvec p^2+M^2}$ is the on-shell energy of the system with mass $M$, is lumped into  the quantity
\begin{equation}
\begin{aligned}
\rho_\Psi(\uvec R,\uvec P)&=\int\ud^3z\,e^{-i\uvec P\cdot\uvec z}\,\Psi^*(\uvec R-\tfrac{\uvec z}{2})\Psi(\uvec R+\tfrac{\uvec z}{2})\\
&=\int\frac{\ud^3q}{(2\pi)^3}\,e^{-i\uvec q\cdot\uvec R}\,\tilde\Psi^*(\uvec P+\tfrac{\uvec q}{2})\tilde\Psi(\uvec P-\tfrac{\uvec q}{2})
\end{aligned}
\end{equation}
which plays the role of a phase-space distribution in the quantum context. The variables $\uvec R$ and $\uvec P$ are interpreted as the average position and momentum of the system. Owing to Heisenberg's uncertainty relations, the quantum phase-space (or Wigner) distribution admits only a quasi-probabilistic interpretation, the usual probabilistic interpretation being recovered under integration over one of the conjugated variables
\begin{equation}
\begin{aligned}
\int\ud^3R\,\rho_\Psi(\uvec R,\uvec P)&=|\tilde\Psi(\uvec P)|^2,\\
\int\frac{\ud^3P}{(2\pi)^3}\,\rho_\Psi(\uvec R,\uvec P)&=|\Psi(\uvec R)|^2.\label{Rprob}
\end{aligned}    
\end{equation}
The matrix element on the RHS of Eq.~\eqref{PSampl} 
\begin{equation}\label{intampl}
\langle O\rangle_{\uvec R,\uvec P}=\int\frac{\ud^3\Delta}{(2\pi)^3}\,e^{i\uvec\Delta\cdot\uvec R}\,\frac{\langle p'|O|p\rangle}{2\sqrt{E_{p'}E_p}}
\end{equation}
with $P=\frac{1}{2}(p'+p)$ and $\Delta=p'-p$, does not depend on the wave packet and can be interpreted as the expectation value in a state localized (in the Wigner sense) around average position $\uvec R$ with average momentum $\uvec P$. 

In the literature, it is often stated that one cannot separate in relativity the center-of-mass motion from the rest of the system. This seems at odds with the phase-space representation in Eq.~\eqref{PSampl}, where the information about the global motion described by $\rho_\Psi(\uvec R,\uvec P)$ is factorized from the internal information contained in $\langle O\rangle_{\uvec R,\uvec P}$. There is in fact no contradiction because $\uvec R$ does not transform as the spatial part of a Lorentz four-vector, contrarily to what is usually implied. It just represents the average position $\uvec R=\frac{1}{2}(\uvec x'+\uvec x)$ of a fictive point defining the center of the system in a given frame (and hence not a material point), where $\uvec x$ and $\uvec x'$ are initial and final eigenvalues of the Newton-Wigner relativistic position operator~\cite{Newton:1949cq,Pavsic:2017orp}. For more details about the problem of defining the center of a relativistic system, see Refs.~\cite{Lorce:2018zpf,Lorce:2021gxs}.

The phase-space representation~\eqref{PSampl} allows one to introduce in a natural way the notion of relativistic 2D spatial distributions for arbitrary average momentum $\uvec P$~\cite{Lorce:2017wkb,Lorce:2018egm}. In this work, we will consider the internal distribution of the charge four-current
\begin{equation}\label{Jampl}
\begin{aligned}
J^\mu_\text{EF}(\uvec b_\perp;P_z)&=\int\ud r_z\,\langle j^\mu(r)\rangle_{\uvec R,P_z\uvec e_z}\\
&=\int\frac{\ud^2\Delta_\perp}{(2\pi)^2}\,e^{-i\uvec\Delta_\perp\cdot\uvec b_\perp}\left[\frac{\langle p',s'|j^\mu(0)|p,s\rangle}{2P^0}\right]_{\Delta_z=|\uvec P_\perp|=0}
\end{aligned}
\end{equation}
with $j^\mu(r)=\sum_f q_f\overline\psi_f(r)\gamma^\mu\psi_f(r)$ the electric charge current operator, $\uvec b_\perp=\uvec r_\perp-\uvec R_\perp$ the impact-parameter coordinates, and the $z$-axis chosen for convenience along the average momentum $\uvec P$. We have included the dependence of the state polarization and integrated over the longitudinal coordinate to ensure that no energy is transferred to the system. In other words, we restricted ourselves to the class of elastic frames (EF) characterized by the condition $\Delta^0=0$~\cite{Lorce:2017wkb}. In the literature one traditionally considers particular cases, namely either the Breit frame $P_z=0$~\cite{Ernst:1960zza,Sachs:1962zzc} or the infinite-momentum frame $P_z\to\infty$ (essentially equivalent to using the light-front formalism)~\cite{Soper:1976jc,Burkardt:2000za,Burkardt:2002hr,Miller:2010nz}. It has been observed that charge distributions in the Breit frame and the infinite-momentum frame can be quite different~\cite{Miller:2007uy,Carlson:2007xd}. Using the interpolating expression in Eq.~\eqref{Jampl}, we showed in Ref.~\cite{Lorce:2020onh} that the distortions can in principle be understood in terms of relativistic kinematical effects associated with spin, and we illustrated this by studying in detail the spin-independent contribution to the charge distribution $J^0_\text{EF}$ inside the nucleon. The spin-dependent contribution has later been worked out in Ref.~\cite{Kim:2021kum}.

In this work, we will study the EF charge distribution~\eqref{Jampl} in the case of a spin-$1$ target and apply our results to the deuteron. Even though the complexity of the analysis increases due to the larger number of allowed multipoles, we will demonstrate that all the distortions can be understood in terms of the same relativistic kinematical effects as those identified in the spin-$1/2$ case.

\section{Spin-1 matrix elements}\label{sec:3}

Matrix elements of the electric charge current for a spin-$1$ target are usually parametrized (in units of the proton electric charge, i.e.~with $e=1$) as~\cite{Arnold:1979cg}
\begin{equation}
\langle p',s'|j^\mu(0)|p,s\rangle=-\varepsilon^*_\alpha(p',s')V^{\mu\alpha\beta}(P,\Delta)\varepsilon_\beta(p,s)
\end{equation}
with
\begin{equation}\label{param}
V^{\mu\alpha\beta}(P,\Delta)=2P^\mu g^{\alpha\beta}\,G_1(Q^2)+(\Delta^\alpha g^{\mu\beta}-\Delta^\beta g^{\mu\alpha})\,G_2(Q^2)-P^\mu\,\frac{\Delta^\alpha\Delta^\beta}{M^2}\,G_3(Q^2)
\end{equation}
and $Q^2=-\Delta^2$. The standard polarization four-vectors are given by
\begin{equation}\label{covpol}
    \varepsilon^\mu(p,s)=\left(\frac{\uvec p\cdot\uvec \epsilon_s}{M},\uvec\epsilon_s+\frac{\uvec p(\uvec p\cdot\uvec\epsilon_s)}{M(p^0+M)}\right)
\end{equation}
with
\begin{equation}\label{restpol}
    \uvec\epsilon_\pm=\tfrac{1}{\sqrt{2}}(\mp 1,-i,0),\qquad \uvec\epsilon_0=(0,0,1)
\end{equation}
representing the rest-frame polarization eigenstates along the $z$-axis. For higher-spin targets, similar parametrizations have been obtained in Refs.~\cite{Lorce:2009bs,Cotogno:2019vjb}.

When expressed in the Breit frame (defined by $\uvec P=\uvec 0$), the matrix elements turn out to exhibit the same spin structure as in the non-relativistic limit~\cite{Arnold:1979cg}
\begin{equation}
\begin{aligned}
    \langle p'_B,s'|j^0(0)|p_B,s\rangle&=2P^0_B\bigg[(\uvec\epsilon^*_{s'}\cdot\uvec\epsilon_s)\,G_C(Q^2)\\
    &\qquad\quad+\left((\uvec\Delta\cdot\uvec\epsilon^*_{s'})(\uvec\Delta\cdot\uvec\epsilon_s)-\tfrac{1}{3}\uvec\Delta^2(\uvec\epsilon^*_{s'}\cdot\uvec\epsilon_s)\right)\frac{G_Q(Q^2)}{2M^2}\bigg],\\
     \langle p'_B,s'|\uvec j(0)|p_B,s\rangle&=2P^0_B\,\big[(\uvec\Delta\cdot\uvec\epsilon^*_{s'})\uvec\epsilon_s-(\uvec\Delta\cdot\uvec\epsilon_s)\uvec\epsilon^*_{s'}\big]\,\frac{G_M(Q^2)}{2M},
\end{aligned}
\end{equation}
where $p^\mu_B=(P^0,-\uvec\Delta/2)$ and $p'^\mu_B=(P^0,\uvec\Delta/2)$ with $P^0_B=\sqrt{M^2+\frac{\uvec\Delta^2}{4}}$. The Breit frame is therefore a natural frame for defining multipole form factors like in the non-relativistic theory. For a spin-1 target, one finds an electric (or Coulomb) monopole $G_C$, a magnetic dipole $G_M$ and an electric quadrupole $G_Q$, given by the combinations
\begin{equation}
    \begin{aligned}
    G_C(Q^2)&=G_1(Q^2)+\tfrac{2}{3}\tau\, G_Q(Q^2),\\
    G_M(Q^2)&=G_2(Q^2),\\
    G_Q(Q^2)&=G_1(Q^2)-G_2(Q^2)+(1+\tau)\,G_3(Q^2),
    \end{aligned}
\end{equation}
where $\tau=Q^2/4M^2$. 

Since the multipole form factors have a clearer physical meaning than those appearing in Eq.~\eqref{param}, it would be interesting to find an alternative parametrization directly in terms of $G_C$, $G_M$ and $G_Q$ displaying in a covariant way the Breit-frame multipole structure. We indeed found that one can equivalently write ($\epsilon_{0123}=+1$)
\begin{equation}\label{newparam}
\begin{aligned}
    V^{\mu\alpha\beta}(P,\Delta)&=2P^\mu\left[\Pi^{\alpha\beta}\,G_C(Q^2)-\frac{\Delta^\rho\Delta^\sigma(\Sigma_{\rho\sigma})^{\alpha\beta}}{2M^2}\,\frac{P^2}{M^2}\,G_Q(Q^2)\right]\\
    &\phantom{=}-\frac{i\epsilon^{\mu\rho\sigma\lambda}\Delta_\rho P_\sigma (\Sigma_\lambda)^{\alpha\beta}}{\sqrt{P^2}}\,G_M(Q^2),
    \end{aligned}
\end{equation}
where we introduced the tensors
\begin{equation}
\begin{aligned}
\Pi^{\alpha\beta}&=g^{\alpha\beta}-\frac{P^\alpha P^\beta}{P^2}+\frac{\Delta^\alpha\Delta^\beta}{4P^2},\\
 (\Sigma^\lambda)^{\alpha\beta}&=-i\epsilon^{\lambda\alpha\beta\omega}\,\frac{P_\omega}{\sqrt{P^2}},\\
  (\Sigma^{\rho\sigma})^{\alpha\beta}&=\frac{1}{2}\,(\Pi^{\rho\alpha}\Pi^{\sigma\beta}+\Pi^{\sigma\alpha}\Pi^{\rho\beta})-\frac{1}{3}\,\Pi^{\rho\sigma}\Pi^{\alpha\beta},
\end{aligned}
\end{equation}
that vanish once contracted with $P_\alpha$ or $P_\beta$ owing to the on-shell constraint $P\cdot\Delta=0$. In the forward limit $\Delta\to 0$, $-\Pi^{\mu\nu}$ reduces to the projector onto the subspace orthogonal to the four-momentum, while $(\Sigma^\lambda)^{\alpha\beta}$ and $(\Sigma^{\rho\sigma})^{\alpha\beta}$ reduce to the vector and tensor polarization operators in the $(\frac{1}{2},\frac{1}{2})$ representation, see e.g. App.~C of Ref.~\cite{Cosyn:2019aio}. Interestingly, Eq.~\eqref{newparam} is reminiscent of the well-known decomposition of the charge current into convection and magnetization currents $\uvec J=\rho\,\uvec v+\uvec\nabla\times\uvec M$~\cite{Yennie:1957}. It follows from discrete spacetime symmetries that the convection (magnetization) contribution can be written as a tower of electric (magnetic) multipoles of even (odd) order. This structure is consistent with the results found for spin-$0$ and spin-$1/2$ targets~\cite{Lorce:2020onh}, and we expect it to apply also to higher-spin targets.

\section{Distortions of the relativistic distributions}\label{sec:4}

We have seen that in the Breit frame (interpreted from the phase-space perspective as the average rest frame of the target), the charge distribution assumes the same spin structure as in the non-relativistic theory. However, as soon as the system starts moving (i.e.~when $\uvec P\neq\uvec 0$) the charge distribution gets distorted by relativistic kinematical effects associated with spin~\cite{Lorce:2020onh}. This can be seen from the general relation~\cite{Jacob:1959at,Durand:1962zza}
\begin{equation}
\langle p',s'|j^\mu(0)|p,s\rangle=\sum_{s'_B,s_B}D^{*(j)}_{s'_Bs'}(p'_B,\Lambda)D^{(j)}_{s_Bs}(p_B,\Lambda)\,\Lambda^\mu_{\phantom{\mu}\nu}\,\langle p'_B,s'_B|j^\nu(0)|p_B,s_B\rangle.
\end{equation}
The first effect of a boost is to mix the Breit-frame charge and current distributions via the Lorentz matrix $\Lambda^\mu_{\phantom{\mu}\nu}$. The second effect is to induce spin rotations via the Wigner rotation matrices $D^{(j)}$ for spin-$j$ targets.

In the Breit frame, the charge (current) distribution defines the electric (magnetic) properties of the system. Under a Lorentz boost, the charge distribution receives therefore a magnetic contribution induced by the global motion of the system. If we ignore the spin rotation effect, the magnetic dipole distribution simply induces a dipolar distortion of the charge distribution when the target is transversely polarized~\cite{Burkardt:2002hr,Carlson:2007xd}. The picture then gets significantly more complicated when spin rotations are included, since they mix the multipole moments. Typically, any individual multipole in the Breit frame will appear as a superposition of all possible multipoles in a boosted frame\footnote{A similar phenomenon is at the origin of model-dependent relations among transverse-momentum dependent parton distributions~\cite{Lorce:2011zta}, and of the link between the pretzelosity distribution and the orbital angular momentum~\cite{Lorce:2011kn}.}. As a result, the boosted charge distribution will contain all multipoles weighted by various combinations of electric and magnetic contributions~\cite{Carlson:2009ovh,Alexandrou:2009hs,Lorce:2009bs,Gorchtein:2009qq,Alexandrou:2012da}. Note however that the total electric charge of the system is Lorentz invariant, and so these relativistic kinematical distortions just reorganize the appearance of the charge distribution in space. 

To illustrate this, let us consider the multipole decomposition of the charge amplitudes in the elastic frame
\begin{equation}
    \left[\frac{\langle p',s'|j^0(0)|p,s\rangle}{2P^0}\right]_{\Delta_z=|\uvec P_\perp|=0}=\mathcal M_0\,\delta_{s's}+\mathcal M^i_1\, (\Sigma^i_0)_{s's}+\mathcal M^{ij}_2\, (\Sigma^{ij}_0)_{s's}+\cdots,
\end{equation}
where $(\Sigma^i_0)_{s's}$ and $(\Sigma^{ij}_0)_{s's}$ are the rest-frame vector and tensor polarization matrices\footnote{The vector polarization matrices are the generalization of the Pauli matrices, and are defined as the spin matrices divided by the spin value.}, and the energy is given by $P^0=\sqrt{M^2(1+\tau)+P^2_z}$. For a spin-$j$ target, the multipole expansion terminates at the $(2j+1)$th term. In the case of a spin-$1/2$ target, the expression for $\mathcal M_0$ has been worked out in Ref.~\cite{Lorce:2020onh}, and the expression for $\mathcal M^i_1$ in Ref.~\cite{Kim:2021kum}. We observe that these results can be put in the simple form
\begin{equation}\label{spinor-multipoles}
\begin{aligned}
    \mathcal M_0&=\frac{M}{P^0}\,G_E(Q^2)+\frac{P^2_z}{P^0(P^0+M)}\,F_1(Q^2),\\
    \mathcal M^i_1&=-\frac{i(\uvec e_z\times\uvec\Delta)^i}{2M}\,\frac{P_z}{P^0}\left[G_M(Q^2)-\frac{P^0}{P^0+M}\,F_1(Q^2)\right],
    \end{aligned}
\end{equation}
where $G_E$ and $G_M$ are the standard Sachs electric and magnetic form factors~\cite{Ernst:1960zza,Sachs:1962zzc}, and
\begin{equation}
    F_1(Q^2)=\frac{1}{1+\tau}\left[G_E(Q^2)+\tau G_M(Q^2)\right]
\end{equation}
is the Dirac form factor. The expression~\eqref{spinor-multipoles} nicely shows that the spin-independent contribution to the charge amplitude is given in the Breit frame (i.e.~when $P_z=0$) by $\mathcal M^B_0=\frac{M}{P^0}\,G_E(Q^2)$ and in the infinite-momentum frame (i.e.~when $P_z\to \infty$) by $\mathcal M^\text{IMF}_0=F_1(Q^2)$. Moreover, it indicates that the magnitude of spin-rotation effects is driven by the sole Dirac form factor. We will see in the next section that a similar structure shows up in the case of spin-$1$ targets.

\section{Spin-$1$ charge amplitudes}\label{sec:5}

For convenience, we define relativistic charge amplitudes as
\begin{equation}
    \mathcal A_{s's}=e^{i(s'-s)\phi_\Delta}\,\left[\frac{\langle p',s'|j^0(0)|p,s\rangle}{2P^0}\right]_{\Delta_z=|\uvec P_\perp|=0} 
\end{equation}
with the momentum transfer written in polar coordinates $\uvec\Delta_\perp=Q\,(\cos\phi_\Delta,\sin\phi_\Delta)$ and the spin quantized along the $z$-axis. For a spin-$1$ target we found\footnote{Our results have recently been confirmed by Ref.~\cite{Kim:2022bia}.} using the parametrization~\eqref{newparam}
\begin{equation}\label{amplres}
    \begin{aligned}
    \mathcal A_{11}&=\mathcal A_{-1-1}=G_C(Q^2)+\frac{\tau}{3}\,G_Q(Q^2)-\frac{\tau P^2_z}{(P^0+M)^2}\,G_W(Q^2,P_z),\\
    \mathcal A_{00}&=G_C(Q^2)-\frac{2\tau}{3}\,G_Q(Q^2)-\frac{2\tau P^2_z}{(P^0+M)^2}\,G_W(Q^2,P_z),\\
    \mathcal A_{01}&=\mathcal A_{-10}=-\frac{\sqrt{\tau}}{\sqrt{2}}\,\frac{P_z}{P^0}\left[G_M(Q^2)+\frac{2P^0[P^0+M(1+\tau)]}{(P^0+M)^2}\,G_W(Q^2,P_z)\right],\\
    \mathcal A_{-11}&=-\tau\,G_Q(Q^2)+\frac{\tau P^2_z}{(P^0+M)^2}\,G_W(Q^2,P_z),
    \end{aligned}
\end{equation}
where
\begin{equation}\label{GW}
\begin{aligned}
    G_W(Q^2,P_z)&=\frac{1}{1+\tau}\left[G_C(Q^2)-\frac{P^0+M(1+\tau)}{P^0}\,G_M(Q^2)+\frac{\tau}{3}\,G_Q(Q^2)\right]\\
    &=G_1(Q^2)-\frac{P^0+M}{P^0}\,G_2(Q^2)+\tau\,G_3(Q^2).    
    \end{aligned}
\end{equation}
The other charge amplitudes are obtained using the relation $\mathcal A_{s's}=(-1)^{s'-s}\mathcal A_{ss'}$. These results can be converted into multipole amplitudes using the polarization matrices given in the Appendix. We found the expressions
\begin{equation}
\begin{aligned}
    \mathcal M_0&=G_C(Q^2)-\frac{4}{3}\,\frac{\tau P^2_z}{(P^0+M)^2}\,G_W(Q^2,P_z),\\
    \mathcal M^i_1&=\frac{i(\uvec e_z\times\uvec\Delta)^i}{2M}\,\frac{P_z}{P^0}\left[G_M(Q^2)+\frac{2P^0[P^0+M(1+\tau)]}{(P^0+M)^2}\,G_W(Q^2,P_z)\right],\\
    \mathcal M_2^{ij}&=-\frac{\Delta^i\Delta^j}{2M^2}\,G_Q(Q^2)-\frac{(\uvec e_z\times\uvec\Delta)^i(\uvec e_z\times\uvec\Delta)^j}{2M^2}\,\frac{P^2_z}{(P^0+M)^2}\,G_W(Q^2,P_z).
    \end{aligned}
\end{equation}
Like in the spin-$1/2$ case, we observe that the magnitude of the contributions arising due to the Wigner spin rotation is driven by a single combination of form factors given in Eq.~\eqref{GW}. A notable difference, however, is that the combination of electromagnetic form factors is now $P_z$-dependent.

The charge amplitudes in Eq.~\eqref{amplres} are defined for any value of the target average momentum. In the limit $P_z\to 0$, we recover the Breit-frame amplitudes with $\Delta_z=0$
\begin{equation}\label{amplBF}
    \begin{aligned}
    \mathcal A^B_{11}&=\mathcal A^B_{-1-1}=G_C(Q^2)+\frac{\tau}{3}\,G_Q(Q^2),\\
    \mathcal A^B_{00}&=G_C(Q^2)-\frac{2\tau}{3}\,G_Q(Q^2),\\
    \mathcal A^B_{01}&=\mathcal A^B_{-10}=0,\\
    \mathcal A^B_{-11}&=-\tau\,G_Q(Q^2).
    \end{aligned}
\end{equation}
In the infinite-momentum frame given by $P_z\to \infty$, our charge amplitudes coincide with those defined within the light-front formalism~\cite{Carlson:2009ovh}
\begin{equation}\label{amplIMF}
    \begin{aligned}
    \mathcal A^\text{IMF}_{11}&=\mathcal A^\text{IMF}_{-1-1}=\frac{1}{1+\tau}\left[G_C(Q^2)+\tau\,G_M(Q^2)+\frac{\tau}{3}\,G_Q(Q^2)\right],\\
    \mathcal A^\text{IMF}_{00}&=\frac{1}{1+\tau}\left[(1-\tau)\,G_C(Q^2)+2\tau\,G_M(Q^2)-\frac{2\tau}{3}\,(1+2\tau)\,G_Q(Q^2)\right],\\
    \mathcal A^\text{IMF}_{01}&=\mathcal A^\text{IMF}_{-10}=-\frac{\sqrt{2\tau}}{1+\tau}\left[G_C(Q^2)-\frac{1}{2}\,(1-\tau)\,G_M(Q^2)+\frac{\tau}{3}\,G_Q(Q^2)\right],\\
    \mathcal A^\text{IMF}_{-11}&=\frac{\tau}{1+\tau}\left[G_C(Q^2)-G_M(Q^2)-\left(1+\frac{2\tau}{3}\right)G_Q(Q^2)\right].
    \end{aligned}
\end{equation}

Parity and time-reversal symmetries imply that all spin-$1$ charge amplitudes can be expressed in terms of $\mathcal A_{11}$, $\mathcal A_{00}$, $\mathcal A_{01}$ and $\mathcal A_{-11}$. Since there are only three electromagnetic form factors allowed by Lorentz symmetry, these four amplitudes must satisfy a linear relation called the angular condition. In the Breit frame with spin quantized along the momentum transfer, one concludes from angular momentum conservation that there cannot be any charge or current amplitude involving a spin-flip of two units~\cite{Grach:1983hd,Keister:1993mg,Capstick:1994ne,Carlson:2003je}. Using the transverse polarization eigenstates given in the Appendix, one finds that
\begin{equation}\label{angcondBF}
    \mathcal A^B_{11}-\mathcal A^B_{00}+\mathcal A^B_{-11}=0
\end{equation}
which is indeed satisfied by the amplitudes in Eq.~\eqref{amplBF}. In terms of light-front or infinite-momentum frame amplitudes, the angular condition becomes~\cite{Grach:1983hd,Keister:1993mg,Capstick:1994ne,Carlson:2003je,Chung:1988my}
\begin{equation}\label{angcondIMF}
    (1+2\tau)\,\mathcal A^\text{IMF}_{11}-\mathcal A^\text{IMF}_{00}+2\sqrt{2\tau}\,\mathcal A^\text{IMF}_{01}+\mathcal A^\text{IMF}_{-11}=0
\end{equation}
as a result of the Melosh-Wigner rotation (i.e.~the Wigner rotation in the limit $P_z\to\infty$). One can easily check that this condition is satisfied by the amplitudes in Eq.~\eqref{amplIMF}. 

Using a generic Wigner rotation matrix for spin-$1$ targets, we found that the angular condition in the class of elastic frames takes the form
\begin{equation}
    (1+2\tan^2\theta)\,\mathcal A_{11}-\mathcal A_{00}-2\sqrt{2}\tan\theta\,\mathcal A_{01}+\mathcal A_{-11}=0,
\end{equation}
and we concluded from the results in Eq.~\eqref{amplres} that the Wigner rotation angle relative to the $z$-axis\footnote{In Ref.~\cite{Carlson:2003je} the angle $\theta$ was defined relative to the negative $x$-axis leading to $\tan\theta=1/\sqrt{\tau}$ in the limit $P_z\to\infty$.} is given by
\begin{equation}
    \tan\theta=-\frac{\sqrt{\tau}\,P_z}{P^0+M(1+\tau)}.
\end{equation}
In the limits $P_z\to 0$ and $P_z\to\infty$, the conditions~\eqref{angcondBF} and~\eqref{angcondIMF} are respectively recovered.

\section{Deuteron relativistic 2D charge distributions}\label{sec:6}

Charge distributions in the elastic frame are obtained by considering the component $\mu=0$ of Eq.~\eqref{Jampl} for a given polarization of the target, and can therefore be expressed in terms of the 2D Fourier transforms of electromagnetic form factors. For the deuteron form factors, we follow the parametrization ``fit II'' of Ref.~\cite{JLABt20:2000qyq} with the parameters given explicitly in Table 1 of Ref.~\cite{Carlson:2009ovh}. 

\begin{figure}[b]
\includegraphics[width=0.4\hsize]{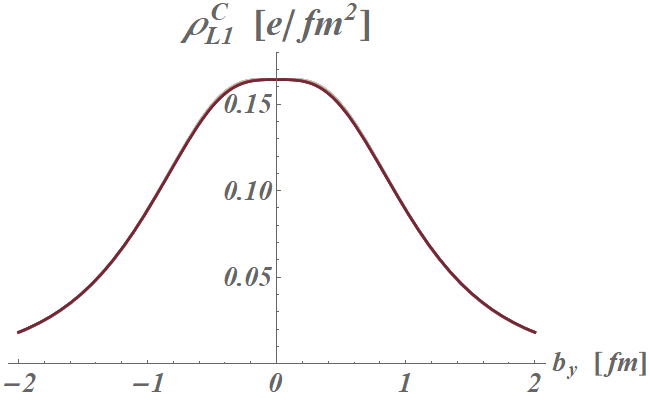}\hspace{.5cm}
\includegraphics[width=0.4\hsize]{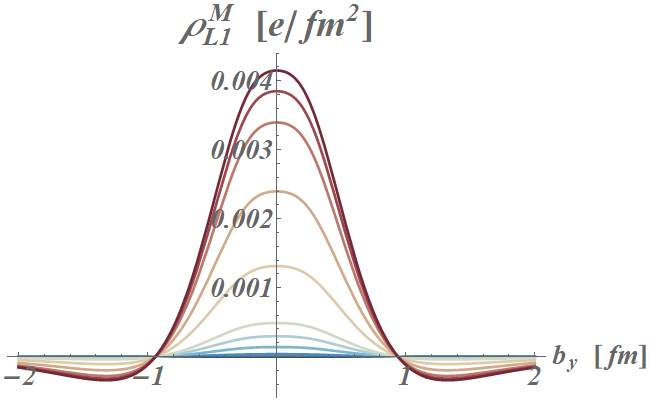}\vspace{.5cm}
\includegraphics[width=0.4\hsize]{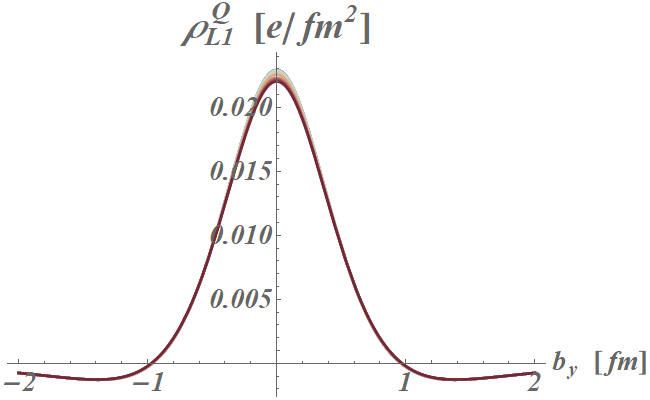}\hspace{.5cm}
\includegraphics[width=0.4\hsize]{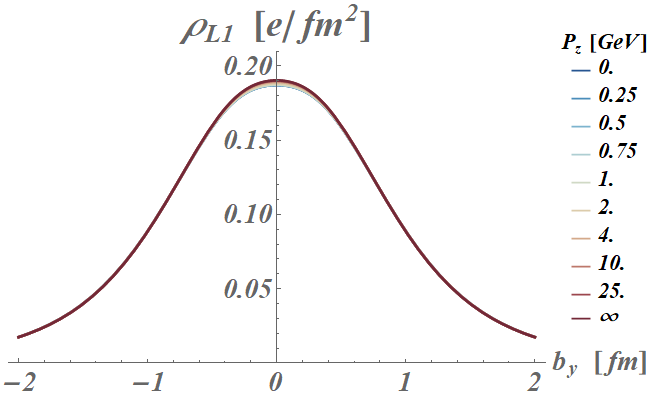}
\caption{Relativistic 2D charge distribution at $b_x=0$ for a deuteron with longitudinal polarization $s_z=1$. From upper-left to bottom-right panel: electric monopole contribution, magnetic dipole contribution, electric quadrupole contribution, and total distribution for selected values of the deuteron average momentum $P_z$. Based on the parametrization ``fit II'' from~\cite{JLABt20:2000qyq}.}\label{fig:rhoL1}
\end{figure}
\begin{figure}[h!]
\includegraphics[width=0.4\hsize]{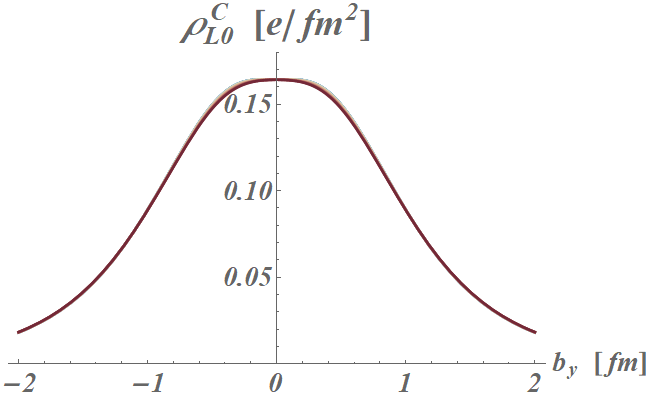}\hspace{.5cm}
\includegraphics[width=0.4\hsize]{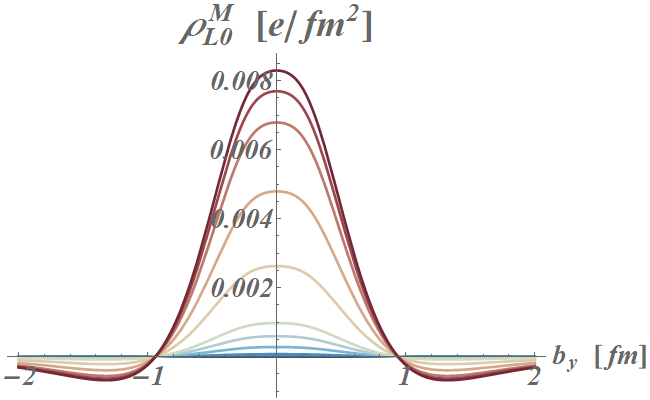}\vspace{.5cm}
\includegraphics[width=0.4\hsize]{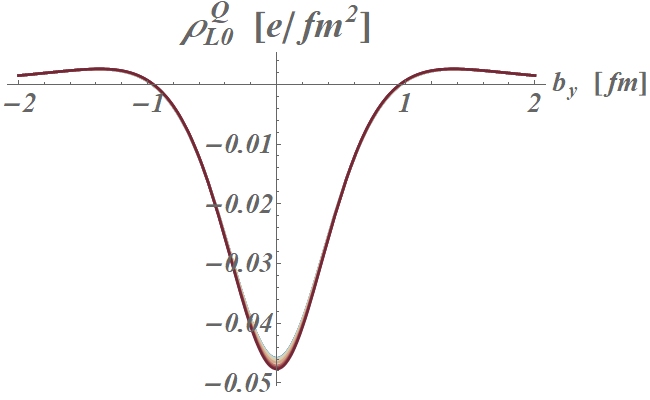}\hspace{.5cm}
\includegraphics[width=0.4\hsize]{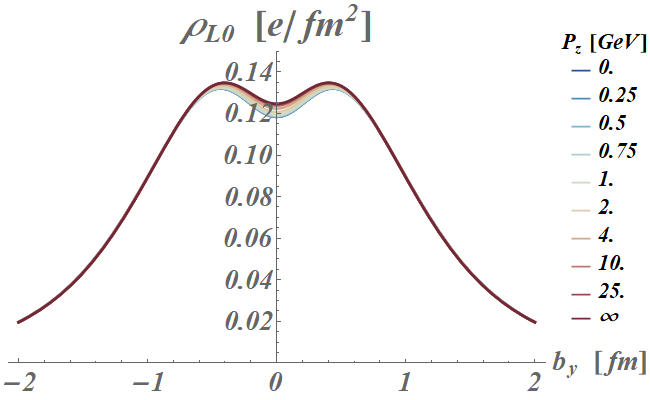}
\caption{Relativistic 2D charge distribution at $b_x=0$ for a deuteron with longitudinal polarization $s_z=0$. From upper-left to bottom-right panel: electric monopole contribution, magnetic dipole contribution, electric quadrupole contribution, and total distribution for selected values of the deuteron average momentum $P_z$. Based on the parametrization ``fit II'' from~\cite{JLABt20:2000qyq}.}\label{fig:rhoL0}
\end{figure}

For a longitudinally polarized deuteron, the 2D charge distributions are axially symmetric and hence simply depend on the radial distance $b=|\uvec b_\perp|$. We can then write
\begin{equation}
       \rho_{L\,s_z}(b;P_z)=\int_0^\infty\frac{\ud Q}{2\pi}\,Q\,J_0(bQ)\,\mathcal A_{s_zs_z},
\end{equation}
where $J_n(bQ)$ is the order-$n$ Bessel function of the first kind. In Figs.~\ref{fig:rhoL1} and~\ref{fig:rhoL0} we show the radial charge distributions with deuteron polarization $s_z=1$ and $s_z=0$ for different values of the average momentum $P_z$. We also provide the individual electric ($\rho^{C,Q}_{L\,s_z}$) and magnetic ($\rho^M_{L\,s_z}$) contributions associated with the corresponding Breit-frame multipole form factors. Like in the nucleon, the electric contributions slightly decrease for increasing values of $P_z$ as a result of the spin rotation. The magnetic contribution (which vanishes by definition at $P_z=0$) increases with $P_z$ but remains here significantly smaller than the electric part. So, contrary to the nucleon case, we do not observe significant $P_z$-dependence in the relativistic 2D charge distributions for a longitudinally polarized deuteron. This can be understood heuristically from the observation that the large relativistic distortions (essentially due to the magnetic contribution) seen in the proton and the neutron charge distributions go in opposite directions, and hence tend to cancel each other in the deuteron case. Another reason is that the deuteron mass and charge radius are both about twice as large as in the nucleon, so that typical values of $\tau$ (which measures relativistic effects) are smaller. 

\begin{figure}[t]
\includegraphics[width=0.4\hsize]{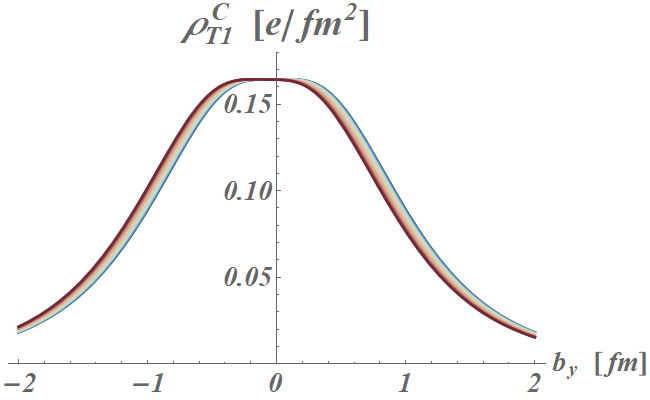}\hspace{.5cm}
\includegraphics[width=0.4\hsize]{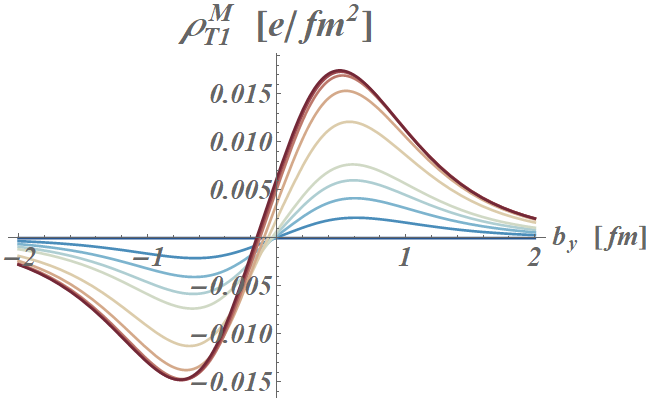}\vspace{.5cm}
\includegraphics[width=0.4\hsize]{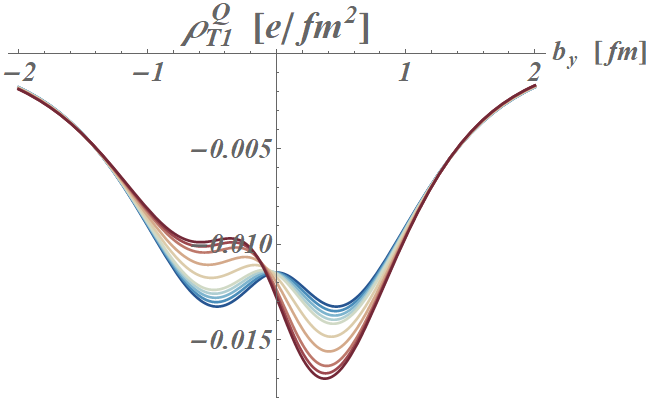}\hspace{.5cm}
\includegraphics[width=0.4\hsize]{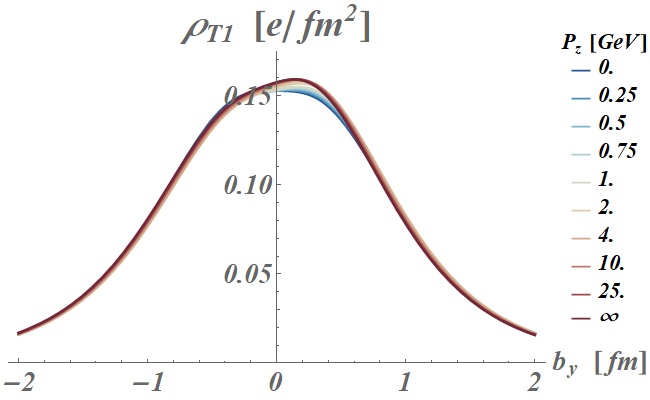}
\caption{Relativistic 2D charge distribution at $b_x=0$ for a deuteron with transverse polarization $s_x=1$. From upper-left to bottom-right panel: electric monopole contribution, magnetic dipole contribution, electric quadrupole contribution, and total distribution for selected values of the deuteron average momentum $P_z$. Based on the parametrization ``fit II'' from~\cite{JLABt20:2000qyq}.}\label{fig:rhoT1}
\end{figure}
\begin{figure}[h!]
\includegraphics[width=0.4\hsize]{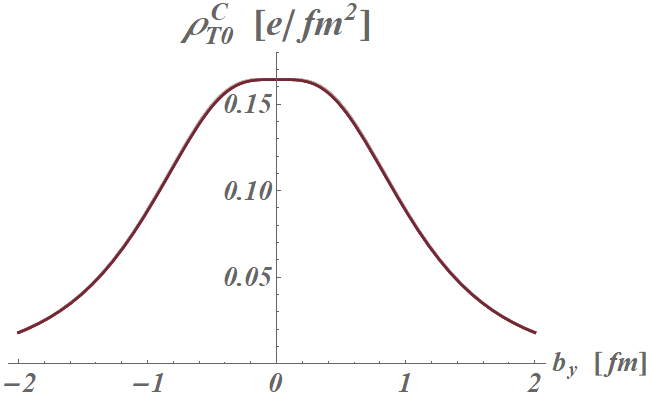}\hspace{.5cm}
\includegraphics[width=0.4\hsize]{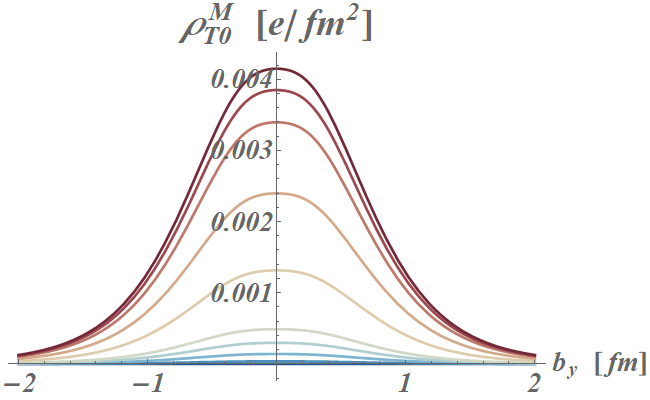}\vspace{.5cm}
\includegraphics[width=0.4\hsize]{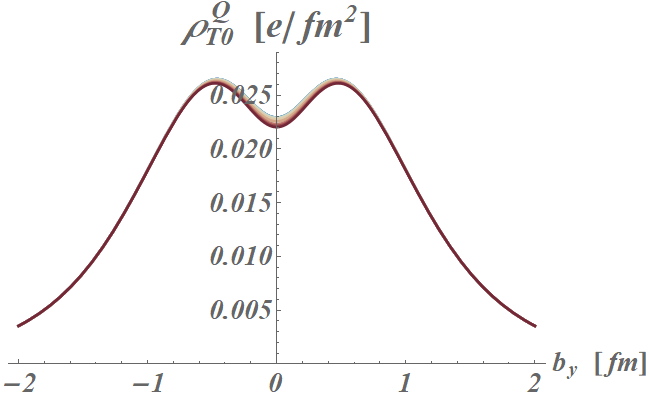}\hspace{.5cm}
\includegraphics[width=0.4\hsize]{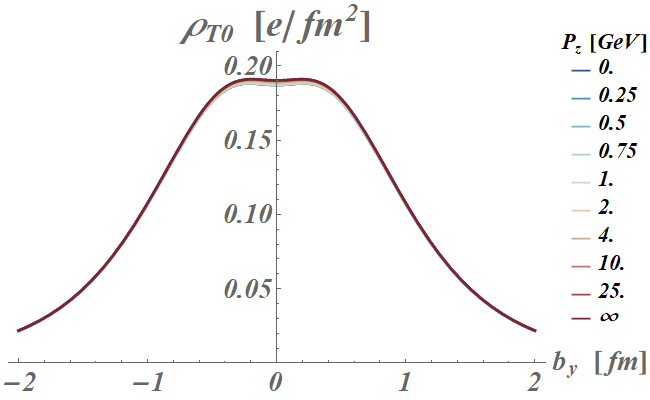}
\caption{Relativistic 2D charge distribution at $b_x=0$ for a deuteron with transverse polarization $s_x=0$. From upper-left to bottom-right panel: electric monopole contribution, magnetic dipole contribution, electric quadrupole contribution, and total distribution for selected values of the deuteron average momentum $P_z$. Based on the parametrization ``fit II'' from~\cite{JLABt20:2000qyq}.}\label{fig:rhoT0}
\end{figure}

When the deuteron is polarized along the transverse direction $\uvec S_\perp=(\cos\phi_S,\sin\phi_S)$, axial symmetry is broken and the relativistic 2D charge distributions are given by
\begin{equation}
    \begin{aligned}
       \rho_{T\,1}(\uvec b_\perp;P_z)&=\int_0^\infty\frac{\ud Q}{2\pi}\,Q\,J_0(bQ)\,\frac{1}{2}\left(\mathcal A_{11}+\mathcal A_{00}\right)\\
       &\phantom{=}+\sin(\phi_b-\phi_S)\int_0^\infty\frac{\ud Q}{2\pi}\,Q\,J_1(bQ)\,\sqrt{2}\,\mathcal A_{01}\\
       &\phantom{=}-\cos2(\phi_b-\phi_S)\int_0^\infty\frac{\ud Q}{2\pi}\,Q\,J_2(bQ)\,\frac{1}{2}\,\mathcal A_{-11},\\
       \rho_{T\,0}(\uvec b_\perp;P_z)&=\int_0^\infty\frac{\ud Q}{2\pi}\,Q\,J_0(bQ)\,\mathcal A_{11}\\
       &\phantom{=}+\cos2(\phi_b-\phi_S)\int_0^\infty\frac{\ud Q}{2\pi}\,Q\,J_2(bQ)\,\mathcal A_{-11},
    \end{aligned}
\end{equation}
using the polarization eigenstates from the Appendix and a polar representation for the relative transverse position $\uvec b_\perp=b\,(\cos\phi_b,\sin\phi_b)$. In Figs.~\ref{fig:rhoT1} and~\ref{fig:rhoT0} we show the charge distributions along the $y$-axis with deuteron polarization $s_x=1$ and $s_x=0$ for different values of the average momentum $P_z$, along with the individual electric ($\rho^{C,Q}_{T\,s_x}$) and magnetic ($\rho^M_{T\,s_x}$) contributions. Similarly to the longitudinally polarized case, we observe a mild $P_z$-dependence. The dipolar distortion is naturally absent for the polarization $s_x=0$ because of parity symmetry. 
\begin{figure}[h!]
\includegraphics[width=0.4\hsize]{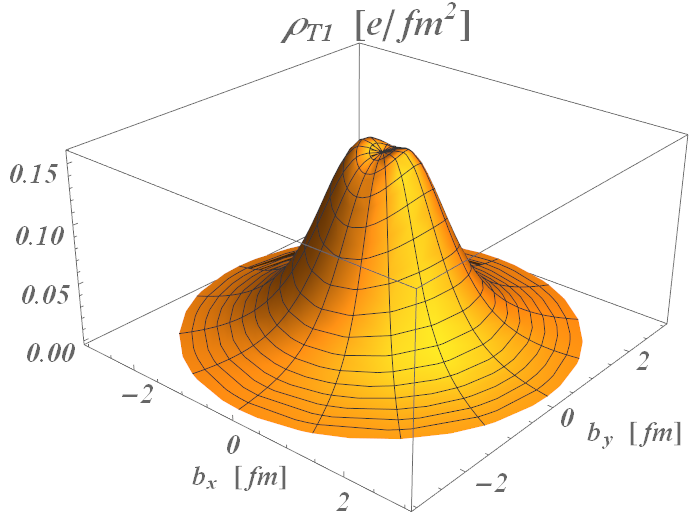}\hspace{.5cm}
\includegraphics[width=0.4\hsize]{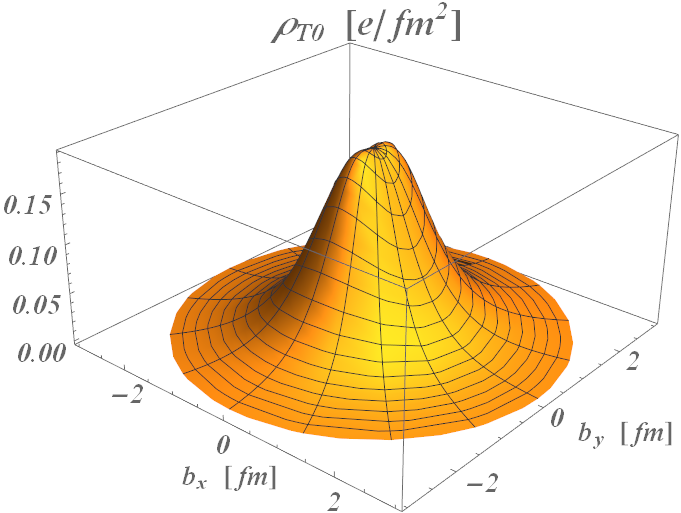}
\caption{Relativistic 2D charge distributions for a deuteron with transverse polarization $s_x=1$ (left panel) and $s_x=0$ (right panel) at the average momentum $P_z=1$ GeV. Based on the parametrization ``fit II'' from~\cite{JLABt20:2000qyq}.}\label{fig:rhoT3D}
\end{figure}
In Fig.~\ref{fig:rhoT3D} we show the relativistic charge distributions in the transverse plane with deuteron polarization $s_x=1$ and $s_x=0$ for $P_z=1$ GeV. The peanut shape is a signature of the large value of the electric quadrupole moment $G_Q(0)=25.84(3)$~\cite{Ericson:1982ei}.

The (induced) electric dipole moment associated with transversely polarized charge distributions is defined as
\begin{equation}
    \uvec d_{\perp\,s_\perp}(P_z)=\int\ud^2 b_\perp\,\uvec b_\perp\,\rho_{T\,s_\perp}(\uvec b_\perp;P_z).
\end{equation}
For a spin-$1$ target, we found (reinstating explicitly the unit of electric charge $e$)
\begin{equation}\label{EDM}
    \uvec d_{\perp\,s_\perp}(P_z)=s_\perp\,(\uvec e_z\times\uvec S_\perp)\,\frac{P_z}{E_P}\left[G_M(0)-\frac{2E_P}{E_P+M}\,G_C(0)\right]\frac{e}{2M},
\end{equation}
where $G_C(0)=1$ and $E_P=\sqrt{P^2_z+M^2}$. Since for the deuteron $G_M(0)=1.71$ and $M=1.875$ GeV~\cite{Tiesinga:2021myr}, the electric dipole moment vanishes and changes sign when $P_z=M\sqrt{\big(\frac{G_M(0)}{2-G_M(0)}\big)^2-1}\approx 11$ GeV. Similarly, the electric quadrupole tensor is defined as
\begin{equation}
    Q^{ij}_{\perp\,s_\perp}(P_z)=\int\ud^2 b_\perp\,(2b^i_\perp b^j_\perp-\delta^{ij}_\perp\,\uvec b^2_\perp)\,\rho_{T\,s_\perp}(\uvec b_\perp;P_z).
\end{equation}
\begin{figure}[t]
\includegraphics[width=0.4\hsize]{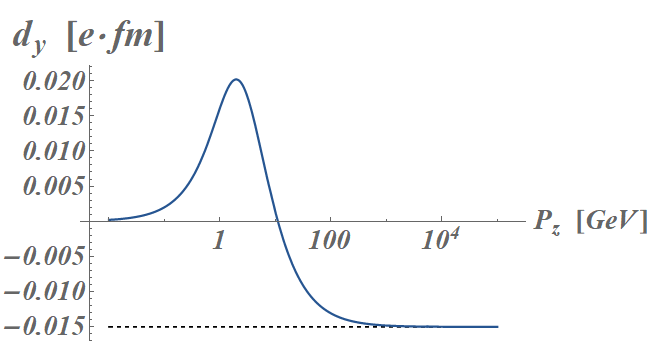}\hspace{.5cm}
\includegraphics[width=0.4\hsize]{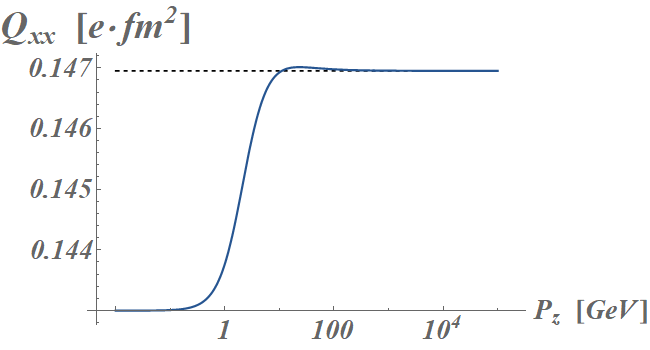}
\caption{Transverse electric dipole (left panel) and quadrupole (right panel) moments as functions of the average momentum $P_z$ for a deuteron with transverse polarization $s_x=1$. Based on the values $G_M(0)=1.71$~\cite{Tiesinga:2021myr} and $G_Q(0)=25.84$~\cite{Ericson:1982ei}.}\label{fig:2Dmultipole}
\end{figure}
For a spin-$1$ target polarized along the $x$-axis, we found
\begin{equation}\label{EQM}
\begin{aligned}
    Q^{xx}_{\perp\,1}(P_z)&=-\frac{1}{2}\,Q^{xx}_{\perp\,0}(P_z)\\
    &=\frac{1}{2}\left\{\left[G_Q(0)+\frac{P^2_z}{(E_P+M)^2}\,G_C(0)\right]\right.\\
    &\phantom{=\frac{1}{2M^2}\quad}\left.+\frac{P^2_z}{E_P(E_P+M)}\left[G_M(0)-\frac{2E_P}{E_P+M}\,G_C(0)\right]\right\}\frac{e}{M^2}.
\end{aligned}
\end{equation}
Once again, these results agree in the limit $P_z\to\infty$ with those obtained within the light-front formalism~\cite{Carlson:2009ovh}. In the limit $P_z\to 0$, the value of the 2D quadrupole is half as large as the 3D one, which is in agreement with the general argument given in Ref.~\cite{Alexandrou:2009hs}. Since $G_Q(0)$ is much larger than $G_C(0)$ and $G_M(0)$, the electric quadrupole moment of the deuteron has a weak $P_z$-dependence. We show in Fig.~\ref{fig:2Dmultipole} the $P_z$-dependence of the transverse electric dipole and quadrupole moments of the deuteron. 

We stress that the results presented in this section are based for illustrative purposes on the parametrization of the experimental data given in Ref.~\cite{JLABt20:2000qyq} two decades ago. As mentioned in the introduction, a more reliable extraction of $G_C(Q^2)$, $G_M(Q^2)$, and $G_Q(Q^2)$ requires a better measurement of the tensor analyzing power in elastic electron-deuteron scattering~\cite{Holt:2012gg}, which should be addressed in a near future by the C1-approved Jefferson Lab experiment C12-15-005~\cite{Long:2015}. Moreover, the future EIC offers a unique and exciting opportunity for measuring these form factors over a larger $Q^2$ range~\cite{AbdulKhalek:2021gbh}, and hence for reducing the uncertainties associated with current extrapolations.

\section{Conclusions}\label{sec:7}

Based on the phase-space formalism, we introduced the relativistic 2D electric charge distributions in a deuteron and studied their dependence on the deuteron average momentum. Contrary to the nucleon case we found a mild frame dependence, reflecting the fact that a deuteron is intrinsically less relativistic than a nucleon. We showed that the two relativistic kinematical effects responsible for the distortions of the charge distributions in a moving frame can nicely be distinguished using a multipole decomposition. Interestingly, we observed that the magnitude of the Wigner spin rotation effects for spin-$1/2$ and spin-$1$ targets is driven by a single combination of electromagnetic form factors. In the limit where the target has infinite momentum, and hence moves almost at the speed of light, all our results reduce to those found earlier within the light-front formalism. We demonstrated once more that the phase-space approach allows one to interpolate between the familiar rest-frame picture (corresponding to the Breit frame) and the light-front picture, where strict probabilistic interpretation is justified.

\appendix

\section*{Appendix: Spin-1 polarization matrices}

Polarization matrices for spin-$1$ targets read
\begin{equation}
\begin{aligned}
    (\Sigma^i_0)_{s's}&=-i\epsilon^{ijk}\epsilon^{*j}_{s'}\epsilon^k_s,\\
(\Sigma^{ij}_0)_{s's}&=\tfrac{1}{2}(\Sigma^i_0\Sigma^j_0+\Sigma^j_0\Sigma^i_0)_{s's}-\tfrac{1}{3}\,\delta^{ij}(\uvec \Sigma_0\cdot\uvec \Sigma_0)_{s's}\\
&=-\tfrac{1}{2}(\epsilon^{*i}_{s'}\epsilon^j_s+\epsilon^{*j}_{s'}\epsilon^i_s)+\tfrac{1}{3}\,\delta^{ij}\delta_{s's},
\end{aligned}
\end{equation}
and are related to the covariant polarization tensors as follows
\begin{equation}
    \begin{aligned}
   \varepsilon^*_\alpha(p_\text{rest},s')(\Sigma^\mu)^{\alpha\beta}\varepsilon_\beta(p_\text{rest},s)&=(0, (\Sigma^i_0)_{s's}),\\
\varepsilon^*_\alpha(p_\text{rest},s')(\Sigma^{\mu\nu})^{\alpha\beta}\varepsilon_\beta(p_\text{rest},s)&=\begin{pmatrix}0&\uvec 0\\\uvec 0&(\Sigma^{ij}_0)_{s's}\end{pmatrix}
    \end{aligned}
\end{equation}
with $p_\text{rest}=(M,\uvec 0)$ the rest four-momentum. Using the polarization vectors~\eqref{restpol}, one finds more explicitly
\begin{equation}
    \Sigma^x_0=\frac{1}{\sqrt{2}}\begin{pmatrix}0&1&0\\ 1&0&1\\ 0&1&0\end{pmatrix},\qquad \Sigma^y_0=\frac{1}{\sqrt{2}}\begin{pmatrix}0&-i&0\\ i&0&-i\\ 0&i&0\end{pmatrix},\qquad
    \Sigma^z_0=\begin{pmatrix}1&0&0\\ 0&0&0\\ 0&0&-1\end{pmatrix},
\end{equation}
and
\begin{equation}
\begin{split}
    \Sigma^{xx}_0=\frac{1}{6}\begin{pmatrix}-1&0&3\\ 0&2&0\\ 3&0&-1\end{pmatrix}&,\qquad \Sigma^{yz}_0=\Sigma^{zy}_0=\frac{1}{2\sqrt{2}}\begin{pmatrix}0&-i&0\\ i&0&i\\ 0&-i&0\end{pmatrix},\\
    \Sigma^{yy}_0=\frac{1}{6}\begin{pmatrix}-1&0&-3\\ 0&2&0\\ -3&0&-1\end{pmatrix}&,\qquad \Sigma^{zx}_0=\Sigma^{xz}_0=\frac{1}{2\sqrt{2}}\begin{pmatrix}0&1&0\\ 1&0&-1\\ 0&-1&0\end{pmatrix},\\
    \Sigma^{zz}_0=\frac{1}{3}\begin{pmatrix}1&0&0\\ 0&-2&0\\ 0&0&1\end{pmatrix}&,\qquad
    \Sigma^{xy}_0=\Sigma^{yx}_0=\frac{1}{2}\begin{pmatrix}0&0&-i\\ 0&0&0\\ i&0&0\end{pmatrix}.
\end{split}    
\end{equation}
Polarization eigenstates along a general direction $\uvec S_\perp=(\cos\phi_S,\sin\phi_S)$ in the transverse plane are given by
\begin{equation}
    |+1\rangle_{\uvec S_\perp}=\frac{1}{2}\begin{pmatrix}1\\ \sqrt{2}\,e^{i\phi_S}\\ e^{2i\phi_S} \end{pmatrix},\quad|0\rangle_{\uvec S_\perp}=\frac{1}{\sqrt{2}}\begin{pmatrix}-1\\ 0\\ e^{2i\phi_S} \end{pmatrix},\quad|-1\rangle_{\uvec S_\perp}=\frac{1}{2}\begin{pmatrix}1\\ -\sqrt{2}\,e^{i\phi_S}\\ e^{2i\phi_S} \end{pmatrix}.
\end{equation}


\end{document}